\documentclass{book}
\usepackage{piers}
\usepackage{cite}

\begin{document}
\title{Phase-sensitive optical time-domain reflectometry with pulse mode EDFA: Probe pulse preparation}
\maketitle

\begin{authors}
{\bf A.O. Chernutsky}$^{1}$, 
{\bf A.A. Zhirnov}$^{1}$, 
{\bf A.K. Fedorov}$^{1,2,3}$, 
{\bf E.T. Nesterov}$^{1}$, \\
{\bf K.V. Stepanov}$^{1}$, 
{\bf Ya.A. Tezadov}$^{4}$, 
{\bf E.V. Kondrashin}$^{4}$,
{\bf V.E. Karasik}$^{1}$
{\bf and A.B. Pnev}$^{1}$\\
\medskip
$^{1}$Bauman Moscow State Technical University, Russia\\
$^{2}$Russian Quantum Center, Russia\\
$^{3}$LPTMS, CNRS, Univ. Paris-Sud, Universit\'e Paris-Saclay, France\\
$^{4}$Scientific and Technological Enterprise IRE-Polyus, Russia
\end{authors}

\begin{paper}

\begin{piersabstract}
Probe pulse preparation techniques play an important role for fiber--optic sensing. 
In this work, we investigate a setup for distributed sensors based on phase-sensitive optical time-domain reflectometry ($\Phi$-OTDR) 
with the use of an erbium doped fibre amplifier (EDFA).
In our $\Phi$-OTDR setup with pulse mode EDFA, 
we reveal a regime providing us both sufficiently large amplification of the probe pulse power together and adequately small probe pulse distortion.
We then show that the use of pulse mode EDFA is a cost-effective solution, 
which allows one to avoid nonlinear effect and ensure sufficient sensitivity of the setup.
\end{piersabstract}

\psection{Introduction}

During last decades, significant attention has been paid to distributed fiber optic sensing systems, 
which are of great importance for monitor of extended objects~\cite{Bao,Shi,Taylor,Taylor2,Taylor3,Taylor4,Taylor5,Kulchin,Fedorov,Sun,Duan,Zhang,Liang,Liang2}.
One of the most promising sensing technique is OTDR~\cite{Bao}.
Standard OTDR uses light sources with coherence lengths, which are much shorter than pulse lengths~\cite{Gold}. 
This system can yield a sum of backscattered intensities from each scattering center, which allows one to reveal splices and breaks in fiber optic cables. 

$\Phi$-OTDR-based sensors work in another regime~\cite{Taylor,Taylor2,Taylor3,Taylor4}. 
In this case, the coherence length of lasers is longer than their pulse length. 
Such systems can be used for real-time vibration control of extended objects, {\it e.g.}, bridges, pipelines, national borders, and many others~\cite{Bao,Shi}. 
Events near the fiber generate an acoustic wave.
This wave changes the phases of the backscattering centers.
An analysis of such signals can reveal their impact on the sensor and monitor located near fiber objects.
An important issue towards to practical implementation of such systems is a development of event recognition algorithms~\cite{Sun,Duan,Zhang,Fedorov2,Fedorov3,Makarenko}. 

A technique for probe pulse preparation is an important stage of operating of $\Phi$-OTDR-based sensors. 
This technique is aimed on both providing sufficient power of the signal and satisfying requirements on the probe pulses shape. 
In order to increase sensitivity of such sensing devices, highly coherent narrowband lasers are typically used~\cite{Shi}.
However, this leads to a significant increase of the cost of such sensing devices. 
Thus, optimization of probe pulse preparation technique is an important step in development of $\Phi$-OTDR-based sensing devices~\cite{Pasquale}.

Practical applications of $\Phi$-OTDR-based sensors require sufficiently high power of source and large coherence length.
Another important characteristic is a shape of probe pulses, which should be close to the rectangular shape. 
It should be noted that these requirements could not be satisfied by using proper semiconductor lasers only. 
A typical way to achieve this is to use following tools: 
narrow linewidth laser with large coherence length, 
EDFA with high output power of signals, 
and acousto-optic modulator (AOM) for providing proper pulse shape~\cite{Shi}.

At this moment, 
both schemes with EDFA precedes AOM ({\it i.e.}, continuous mode EDFA, see Refs.~\cite{Gonzalez-Herraez,Gonzalez-Herraez2,Zhang2}) 
and schemes with their inverse sequence (pulse mode EDFA, see Refs.~\cite{Izumita,Rao,Zeng,Hartog,Li}) are in use. 
We note that the pulse mode EDFA scheme makes requirements to AOM weaker~\cite{Izumita,Li,Rao,Hartog,Zeng}, 
whereas EDFA in the continuous mode should be accomplishment by a special stabilization scheme giving rise to $0.1$ dB stability level~\cite{Wen}.

One of the key challenge of using of pulse mode EDFA schemes is as follows. 
In this case the pulse shape is determined not only by AOM, but also transients in EDFA.
In its turn, this leads to a distortion of the shape of probe pulses.
Therefore, even if the input power is lower that the critical value~\cite{Pnev}, the pulse distortion leads to possible appearance of nonlinear effects. 
Thus, an interesting and important task is to find a regime providing both sufficiently large amplification of probe pulse power and adequately small gradient of pulse shapes 
(not very large distortion) for pulse mode EDFA setups for $\Phi$-OTDR-based sensors.
This could provide a cost-effective solution, which allows one pushing frontiers of applications of $\Phi$-OTDR-based sensors forward and explore new markets.

In this work, we demonstrate experimental results on the use of a low-cost narrowband (${<}100$ kHz) 
semiconductor laser diode with pulse mode EDFA for probe pulses preparation (see Fig.~\ref{fig:setup}).
We show that nonlinear effects could appear due to the distortion of shape of probe pulses even below the critical value of power for our setup~\cite{Pnev}.
This is realized by measurements of pulse shape dependences on pump power, pulse duration, and pulse repetition frequency.
We then find a regime, which allows avoiding nonlinear effect and providing sufficient sensibility.

Our paper is organized as follows.
In Sec.~\ref{sec:experiment}, we study influences of nonlinear effects due to the pulse shape distortion on operating of the $\Phi$-OTDR-based sensor.
In Sec.~\ref{sec:numerics}, we implement a numerical analysis of influences of the probe pulse shape on resulting data from our $\Phi$-OTDR-based sensing setup. 
We conclude and summarize our results in Sec.~\ref{sec:conclusion}. 

\psection{Experimental study}\label{sec:experiment}

\begin{figure}[t]
\begin{centering}
\includegraphics[width=0.75\columnwidth]{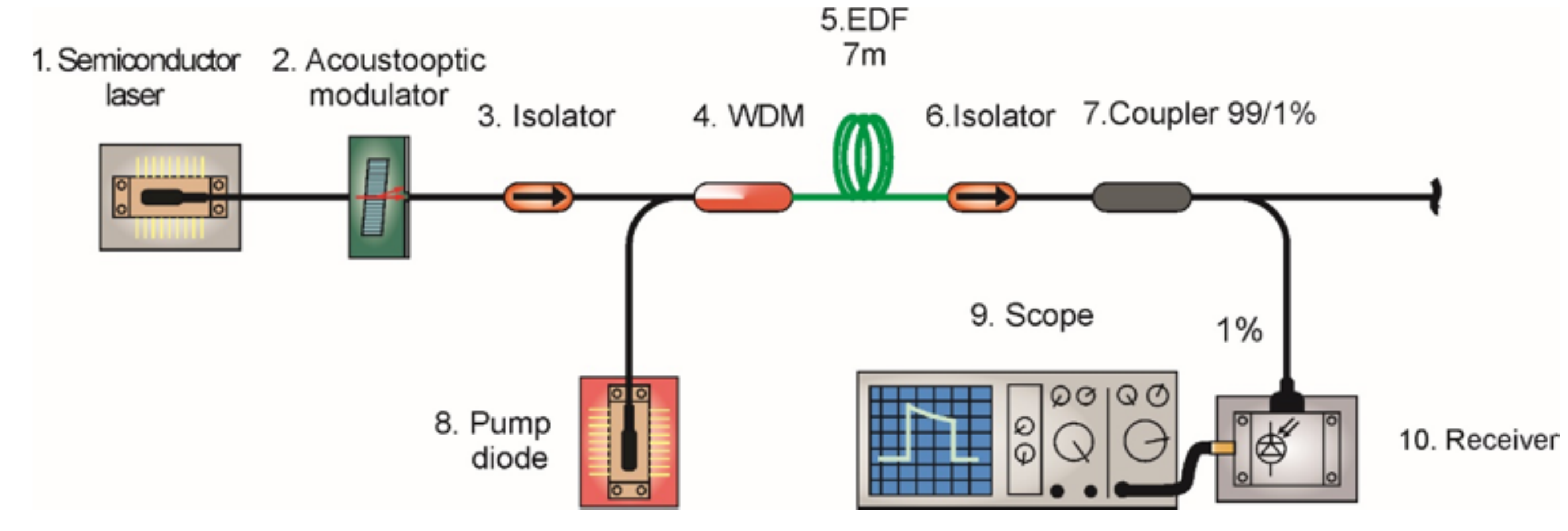}
\vskip -4mm
\caption
{Suggested setup for $\Phi$-OTDR-based sensor with pulse mode EDFA: signal from a continuous-wave laser enters to the AOM. 
The resulting pulse of output power is launched into a fiber.}
\label{fig:setup}
\end{centering}
\end{figure}

Our setup for investigation of influences of nonlinear effects due to the pulse shape distortion on operating of the $\Phi$-OTDR-based sensor is presented in Fig.~\ref{fig:setup}.
A signal from the continuous-wave laser (wavelength is $\lambda{=}1550$ nm and output power is about 10 mW) enters to the AOM. 
The resulting pulse is launched into the EDF HP980. 
The length $L$ is about $7$ meters. 
In our setup, the PIN photodiode with known sensitivity at the signal wavelength is used as the receiver. 
The resulting data are reflectograms, {\it i.e.}, space-time intensity distributions. 

We note that under certain conditions the leading edge of the modulated optical pulse in EDFA depletes the inversion in an erbium fiber medium. 
This results in a higher gain than the trailing edge~\cite{Pnev}. 
Therefore, the waveform of the signal pulse is deformed during its transmission through the amplifier. 
In fact, this leads to the self-phase modulation due to frequency shift of backscattered signal~\cite{Pnev}. 

It is rather well known that the self-phase modulation appears due to the power dependence of the refractive index in a quartz medium of a single mode optical fiber. 
The higher intensity of an optical pulse trailing edge encounter a higher refractive index of the fiber compared with the lower intensity of leading edge, 
while optical probe pulse passing through the optical fiber. 
In this case, the leading edge have a positive gradient of refractive index ($dn/dt$), while the trailing edge a negative gradient of refractive index ($-dn/dt$).
Then the gradient of pulse power produces a time varying refractive index in a medium, which leads to time varying phase change. 
Different parts of the pulse undergo different phase shift.
Thus, the optical frequency of the transmitted optical pulse are exposed to a frequency chirp. 

One can calculate the critical value of the gradient of probe pulse shapes as follows. 
The optical frequency shift $\delta\nu$ can be presented in the following form~\cite{Izumita}:
\begin{equation}\label{eq:1}
	\delta\nu=\frac{n_2L_{\rm eff}}{A\lambda}\frac{\partial{p(t)}}{\partial{t}}.
\end{equation}
Here, $n_2$ is the nonlinear refractive index, and the effective length $L_{\rm eff}$ has the form:
\begin{equation}
	L_{\rm eff}=\frac{1-\exp({-\alpha{L}})}{\alpha},
\end{equation}
where 
$A=\pi{a_m}^2$ is the effective cross section area, 
$a_m$ is the mode-field radius, 
$\lambda$ is the incident optical pulse wavelength,
$p(t)$ is the dependence of an optical pulse power on time $t$, 
$L$ is the length of the optical fiber,
and 
$\alpha$ is the attenuation coefficient.
Using parameters of our setup and Eq. (\ref{eq:1}), we obtain that the critical power gradient of optical probe pulse is $\sim{4}\times10^{5}$.

We start our experimental investigation from measurements of the dependence of averaging output power of the pulse on pump power
at fixed pulse duration ($\tau=200$ ns) and fixed pulse repetition frequency ($F{\sim}15$ kHz). 
In order to obtain accurate results, we use the 99/1\% coupler to avoid the saturation mode of the receiver. 

\begin{figure*}
\includegraphics[width=1\columnwidth]{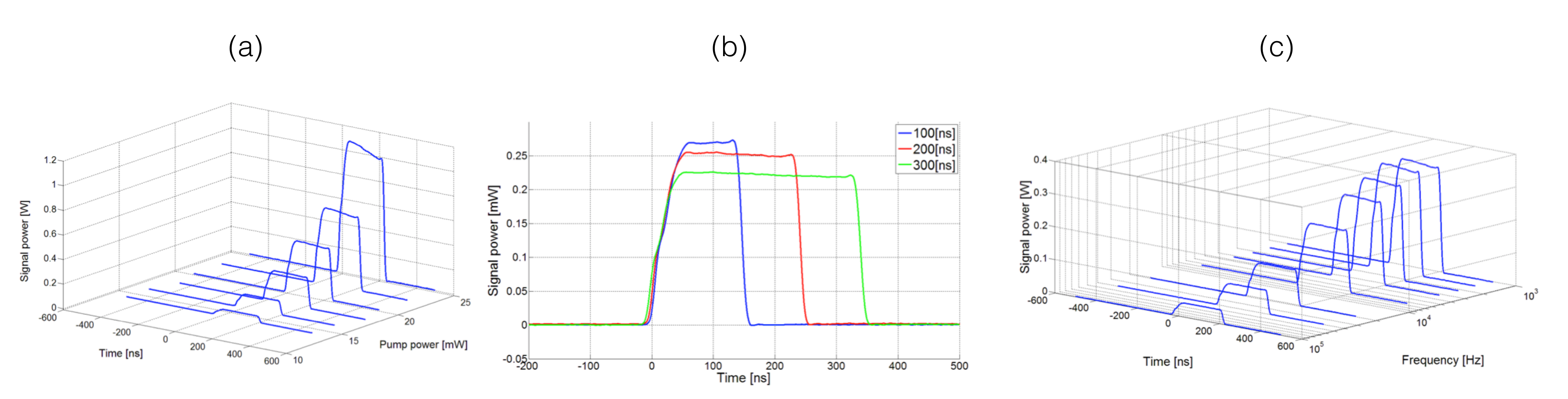}
\vskip -2mm
\caption
{
Experimental study of the role of nonlinear effects due to the probe pulse distortion on operating of $\Phi$-OTDR-based fiber sensor with pulse mode EDFA. 
In (a) the average output (signal) power of the pulse is shown as a function of the pump power and time at fixed pulse duration ($\tau=200$ ns) and pulse repetition frequency ($F{\sim}15$ kHz).
In (b) the output power level is shows as a function of time for three different pulse durations: $\tau_1=100$ ns (blue curve), $\tau_2=200$ ns (red curve), and $\tau_3=300$ ns (green curve).
In (c) the signal power is shown as a function of time and the pulse repetition frequency.
}
\label{fig:nonlinear}
\end{figure*}

The experimental results are presented in Fig.~\ref{fig:nonlinear}a.
One can see that the pulse shape is deformed significantly at increasing of the pump power in erbium doped fiber. 

We then demonstrate the dependence of output power level on time for three different pulse durations: $\tau_1=100$ ns, $\tau_2=200$ ns, and $\tau_3=300$ ns
at the constant power of the pump (20 mW) in Fig.~\ref{fig:nonlinear}b.
The distortion value of corresponding pulse-shapes comparing to perfectly rectangle pulse-shape was 0.3\%, 2.1\%, and 3\% for each pulse duration, respectively.
On this step, one can calculate the values of the gradient for these three probe pulses. 
They are on the level $8\times{10}^3$, $3.5\times{10}^4$, and $3\times{10}^4$.
These values are below the critical value for our setup ($3.76\times10^{5}$), which is given by Eq. (\ref{eq:1}).
It is important to study the dependence of the pulse shape and power level on the pulse repetition frequency as well \cite{Zhihong}. 
The experimental results on measurement of this dependence is shown in Fig.~\ref{fig:nonlinear}c. 
In this experiment, the duration of pulse ($\tau=200$ ns) and pump power in EDFA ($P=20$ mW) are constant, 
the pulse repetition frequency changes from 10 up to 100 kHz.
This frequency band is typical for operating of $\Phi$-OTDR-based sensors. 

\psection{Numerical analysis}\label{sec:numerics} 

\begin{figure*}
\includegraphics[width=1\columnwidth]{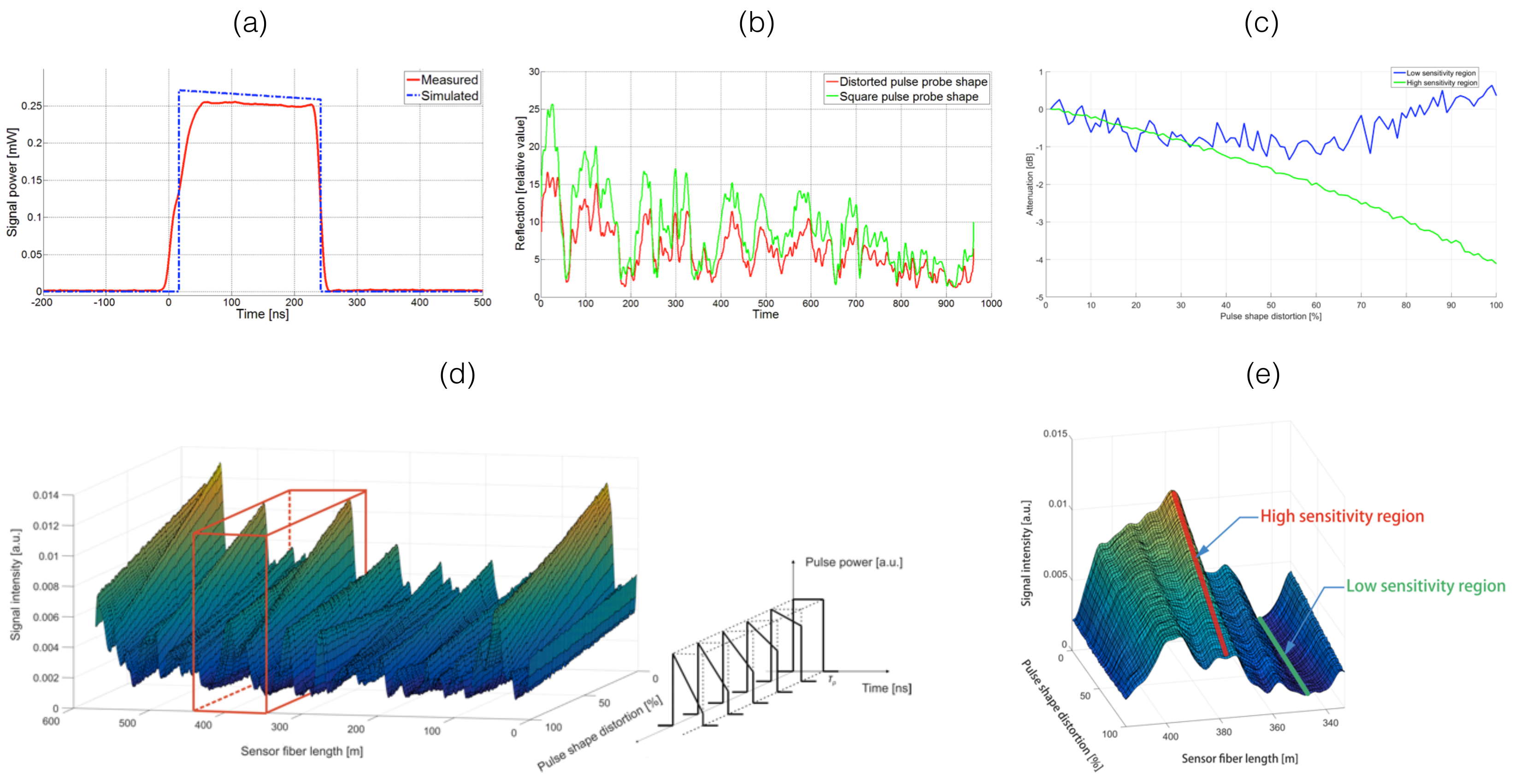}
\caption
{
Result of the numerical analysis.
In (a) the simulated shape of the distorted probe pulse (blue) and the measured shape probe pulse (red) are shown.
In (b) the simulated resulting data from the sensing system after probing by distorted (green) and perfectly rectangle (red) probe pulses.
We note that the resulting reflectograms are quite similar. 
In (c) the decrease of the intensity of the probe pulse is shown as function of the value of the pulse shape distortion. 
We note that the decrease of the intensity [see also Eq.~(\ref{eq:decrease})] is on the level of 4 dB at the maximum value of the distortion of probe pulses. 
In (d) the intensity of the signal is shown as a function of the fiber length and the pulse shape distortion.
In (e) the same dependency with highlighted regions of high sensitivity and low sensitivity. 
}
\label{fig:data}
\end{figure*}

Here we study a dependence of the probe pulse shape on resulting data (reflectograms) from our $\Phi$-OTDR-based sensing setup. 
We use numerical analysis of receiving reflection data with two different type of optical pulse shapes: perfectly rectangle and distorted. 
The numerical analysis is based on the mathematical model suggested in Refs.~\cite{Pnev2,Pnev3,Cao}.
For numerical simulation of probe pulses, we also employ the simplified Franz-Nodvik mathematical model~\cite{Frantz}.

We use numerical simulation of reflectograms with varying the shape of probe pulses at fixed other parameters (scattering center positions, amplifiers noises, laser wavelength and etc).
The results are presented in Fig~\ref{fig:data}. 
The shape changes in 100 steps from rectangular to triangular as shown in Fig. \ref{fig:data}d. 
We highlight high (peak) and low (bottom) sensitivity regions in Fig.~\ref{fig:data}e. 
For these two regions, we calculate the decrease of the intensity of probe pulses as follows: 
\begin{equation}\label{eq:decrease}
	I_s(\mu)=10\log\left(\frac{I(\mu)}{I_{\rm max}}\right),
\end{equation}
where 
$I(\mu)$ is the backscattering signal,
the parameter $\mu$ describes the distortion of pulses ({\it i.e.}, $\mu=0$ for a pulse of the rectangular form),
and $I_{\rm max}$ is the value of the backscattering signal ({\it i.e.}, $I_{\rm max}=I(0)$ at $\mu=0$).

The form of probe pulse plays a crucial role for recognition procedures~\cite{Sun,Fedorov2,Fedorov3}. 
One can see that the resulting reflectograms obtained by probing the fiber by distorted and perfectly rectangular pulses
(for comparison of the distorted and perfectly rectangle pulses, see Fig.~\ref{fig:data}b) 
are qualitatively the same, however they are different quantitatively. 
Nevertheless, for obtained experimental distortion (less than 4\%) it is not significant. 
One can also see that the decrease of the intensity in high sensitivity region (4 dB) is much higher than in low sensitivity region. 
Therefore, it is possible to achieve a regime, which allows avoiding of the significant suppression of the intensity and providing the sensitivity for the system.

\psection{Conclusion}\label{sec:conclusion}

We have demonstrated the setup for $\Phi$-OTDR-based sensing systems with pulse mode EDFA. 
Using obtained experimental and simulation results, 
the regime with avoided nonlinear effect and provided sufficient sensitivity has been revealed (see Fig.~\ref{fig:data}c, Fig.~\ref{fig:data}d, and Fig.~\ref{fig:data}e).
Thus, the suggested scheme can be considered as a potential candidate for cost-effective solution for a $\Phi$-OTDR-based distributed sensor as a component 
of multipurpose monitoring systems~\cite{Fedorov4}.

\ack
This work was supported by the Ministry of Education and Science of the Russian Federation under the project 14.577.21.0224 (ID RFMEFI57716X0224).

\end{paper}

\begin{thebibliography}{99}

\bibitem{Bao}
Bao,~X. and Chen,~L.
``Recent progress in distributed fiber optic sensors,"
{\it Sensors}, Vol.~{\bf 12}, 8601, 2012.

\bibitem{Shi}
Shi,~Y., Feng,~H., and Zhoumo,~Z.
``A long distance phase-sensitive optical time domain reflectometer with simple structure and high locating accuracy,"
{\it Sensors}, Vol.~{\bf 15}, 21957, 2015.
 
\bibitem{Taylor}
Taylor,~H.~F. and Lee,~C.~E. 
``Apparatus and method for fiber optic intrusion sensing,"
U.S. Patent 5, 194 847, March 16, 1993.

\bibitem{Taylor2}
Park,~J., Lee,~W., and Taylor,~H.~F.
``Fiber optic intrusion sensor with the configuration of an optical time-domain reflectometer using coherent interference of Rayleigh backscattering,''
{\it Proc. SPIE}, Vol.~{\bf 3555}, 49, 1998.

\bibitem{Taylor3}
Choi,~K.~N. and Taylor,~H.~F.
``Spectrally stable Er-fiber laser for application in phase-sensitive optical time-domain reflectometry,''
{\it IEEE Photon. Technol. Lett.}, Vol.~{\bf 15}, 386, 2003.

\bibitem{Taylor4}
Choi,~N.~K., Juarez,~J.~C., and Taylor,~H.~F.
``Distributed fiber optic pressure/seismic sensor for low-cost monitoring of long perimeters,''
{\it Proc. SPIE}, Vol.~{\bf 5090}, 134, 2003.

\bibitem{Taylor5}
Juarez,~J.~C., Maier,~E.~W., Choi,~K.~N., and Taylor,~H.~F.
``Distributed fiber-optic intrusion sensor system,''
{\it J. Lightwave Technol.}, Vol.~{\bf 23}, 2081, 2005.

\bibitem{Kulchin}
Kulchin,~Yu.~N., Vitrik,~O.~B., Dyshlyuk,~A.~V., Shalagin,~A.~M., Babin,~S.~A., and Vlasov,~A.~A.
``Application of optical time-domain reflectometry for the interrogation of fiber Bragg sensors,''
{\it Laser Phys.}, Vol.~{\bf 17}, 1335, 2007.
 
\bibitem{Fedorov}
Fedorov,~A.~K., Lazarev,~V.~A., Makhrov,~I.~P., Pozhar,~N.~O., Anufriev,~M.~N., Pniov,~A.~B., and Karasik,~V.~E.
``Structural monitoring system with fiber Bragg grating sensors: implementation and software solution,''
{\it J. Phys.: Conf. Ser.}, Vol.~{\bf 594}, 012049, 2015.

\bibitem{Sun}
Sun,~Q., Feng,~H., Yan,~X., and Zeng,~Z.,
``Recognition of a phase-sensitivity OTDR sensing system based on morphologic feature extraction,''
{\it Sensors}, Vol.~{\bf 15}, 15179, 2015.

\bibitem{Duan}
Duan,~N., Peng,~F., Rao,~Y.-J., Du.~J., and Lin.~Y.
``Field test for real-time position and speed monitoring of trains using phase-sensitive optical time domain reflectometry ($\Phi$-OTDR),''
{\it Proc. SPIE}, Vol.~{\bf 9157}, 91577A, 2014.

\bibitem{Zhang}
Qu,~H., Ren,~X., Li,~G., Li,~Y., and Zhang,~C.
``Study on the algorithm of vibration source identification based on the optical fiber vibration pre-warning system,''
{\it Photonic Sens.}, Vol.~{\bf 5}, 180, 2015.

\bibitem{Liang}
Liang,~S., Sheng,~X., Lou,~S., Feng,~Y., and Zhang,~K.
``Combination of phase-sensitive OTDR and Michelson interferometer for nuisance alarm rate reducing and event identification,''
{\it IEEE Photonics J.}, Vol.~{\bf 6}, 6802112, 2016.
 
 \bibitem{Liang2}
Liang,~S., Sheng,~X., and Lou,~S.
``Experimental investigation on lower nuisance alarm rate phase-sensitive OTDR using the combination of a Mach--Zehnder interferometer,''
{\it Infrared Phys. Tech.}, Vol.~{\bf 75}, 117, 2016.

\bibitem{Gold}
Gold,~M.~P.
``Design of a long-range single-mode OTDR,''
{\it J. Lightwave Technol.}, Vol.~{\bf 3}, 39, 1985.

\bibitem{Fedorov2}
Fedorov,~A.~K., Anufriev,~M.~N., Zhirnov,~A.~A., Nesterov,~E.~T., Namiot,~D.~E., Pnev,~A.~B., and Karasik,~V.~E.
``Towards events recognition in a distributed fiber-optic sensor system: Kolmogorov-Zurbenko filtering,''
{\it Int. J. Open Inform. Techn.}, Vol.~{\bf 3}, 16, 2015.

\bibitem{Fedorov3}
Fedorov,~A.~K., Anufriev,~M.~N., Zhirnov,~A.~A., Stepanov,~K.~V., Nesterov,~E.~T., Namiot,~D.~E., Karasik,~V.~E., and Pnev,~A.~B.
``Gaussian mixture model for event recognition in optical time-domain reflectometry based sensing systems,''
{\it Rev. Sci. Instrum.}, Vol.~{\bf 87}, 036107, 2016.

\bibitem{Makarenko}
Makarenko,~A.~V.
``Deep learning algorithms for signal recognition in long perimeter monitoring distributed fiber optic sensors,''
{\it Proceedings of the IEEE 26th International Workshop on Machine Learning for Signal Processing (Vietri sul Mare, Italy, 2016)} p. 1.

\bibitem{Pasquale}
Muanenda,~Y., Oton,~C.~J., Faralli,~S., and Di Pasquale,~F.
``A cost-effective distributed acoustic sensor using a commercial off-the-shelf DFB laser and direct detection phase-OTDR,''
{\it IEEE Photonics J.}, Vol.~{\bf 8}, 6800210, 2016.

\bibitem{Gonzalez-Herraez}
Martins,~H.~F., Mart\'in-L\'opez,~S., Corredera,~P., Filograno,~M.~L., Fraza\~o,~O., and Gonzalez-Herr\'aez,~M.
``Phase-sensitive optical time domain reflectometer assisted by first-order Raman amplification for distributed vibration sensing over $>100$ km,''
{\it J. Lightwave Technol.}, Vol.~{\bf 32}, 1510, 2014.

\bibitem{Gonzalez-Herraez2}
Martins,~H.~F., Mart\'in-L\'opez,~S., Corredera,~P., Ania-Casta\~non,~J.~D., Fraza\~o,~O., and Gonzalez-Herr\'aez,~M.
``Distributed vibration sensing over 125 km with enhanced SNR using Phi-OTDR over a URFL cavity,''
{\it J. Lightwave Technol.}, Vol.~{\bf 33}, 2628, 2015.

\bibitem{Zhang2}
Zhou,~L., Wang,~F., Wang,~X., Pan,~Y., Sun,~Z., Hua,~J., and Zhang,~X.
``Distributed strain and vibration sensing system based on phase-sensitive OTDR,''
{\it IEEE Photonics Technol. Lett.}, Vol.~{\bf 27}, 1884, 2015.

\bibitem{Izumita}
Izumita,~H., Koyamada,~Y., Furukawa,~S., and Sankawa,~I.
``The performance limit of coherent OTDR enhanced with optical fiber amplifiers due to optical nonlinear phenomena,"
{\it J. Lightwave Technol.}, Vol.~{\bf 12}, 1230, 1994.

\bibitem{Li}
Fang,~G., Xu,~T., Feng,~S., and Li,~F.
``Phase-sensitive optical time domain reflectometer based on phase-generated carrier algorithm,"
{\it J. Lightwave Technol.}, Vol.~{\bf 33}, 2811, 2015.

\bibitem{Rao}
Wu,~H., Xiao,~S., Li,~X., Wang,~Z., Xu,~J., and Rao,~Y.
``Separation and determination of the disturbing signals in phase-sensitive optical time domain reflectometry ($\Phi$-OTDR),"
{\it J. Lightwave Technol.}, Vol.~{\bf 33}, 3156, 2015.

\bibitem{Hartog}
Liokumovich,~L.~B., Ushakov,~N.~A., Kotov,~O.~I., Bisyarin,~M.~A., and Hartog,~A.~H.
``Fundamentals of optical fiber sensing schemes based on coherent optical time domain reflectometry: signal model under static fiber conditions,"
{\it J. Lightwave Technol.}, Vol.~{\bf 33}, 3660, 2015.

\bibitem{Zeng}
Shi,~Y., Feng,~H., and Zeng,~Z.
``Phase-sensitive optical time domain reflectometer with dual-wavelength probe pulse,"
{\it Int. J. Distrib. Sens. N.}, Vol.~{\bf 11}, 624643, 2015.

\bibitem{Wen}
Li,~H., Zhang,~Y., Soh,~Y.~C., and Wen,~C.
``Design and analysis of dynamic erbium-doped fiber amplifier gain-clamping systems with feedback control,"
{\it J. Opt. Soc. Am. B}, Vol.~{\bf 24}, 1739, 2007.

\bibitem{Pnev}
Nesterov,~E.~T., Zhirnov,~A.~A., Stepanov,~K.~V., Pnev,~A.~B., Karasik,~V.~E., Tezadov,~Ya.~A., Kondrashin,~E.~V, and Ushakov,~A.~B.
``Experimental study of influence of nonlinear effects on phase-sensitive optical time-domain reflectometer operating range,"
{\it J. Phys.: Conf. Ser.}, Vol.~{\bf 584}, 012028, 2015.

\bibitem{Zhihong}
Li,~Z., Heidt,~A.~M., Tec,~P.~S., Berendt,~M., Sahu,~J.~K., Phelan,~R., Kelly,~B., Alam,~S.~U., and Richardson,~D.~J.
``High-energy diode-seeded nanosecond 2$\mu$m fiber MOPA systems incorporating active pulse shaping,"
{\it Opt. Lett.}, Vol.~{\bf 39}, 1569, 2014.

\bibitem{Pnev2}
Pnev,~A.~B., Zhirnov,~A.~A., Stepanov,~K.~V., Nesterov,~E.~T., Shelestov,~D.~A., and Karasik,~V.~E.
``Mathematical analysis of marine pipeline leakage monitoring system based on coherent OTDR with improved sensor length and sampling frequency,''
{\it J. Phys.: Conf. Ser.}, Vol.~{\bf 584}, 012016, 2015.

\bibitem{Pnev3}
Zhirnov,~A.~A., Fedorov,~A.~K., Stepanov,~K.~V., Nesterov,~E.~T., Chernutsky,~A.~O., Sazonkin,~S.~G., Shelestov,~D.~A., Karasik,~V.~E., Svelto,~C., and Pnev,~A.~B.
``Effects of laser frequency drift in phase-sensitive optical time-domain reflectometry fiber sensors,"
{\it arXiv.org:1604.08854}.

\bibitem{Cao}
Zhong,~X., Zhang,~C., Li, Liang,~S., Li,~Q., L\"u,~Q., Ding,~X., and Cao,~Q.
``Influences of laser source on phase-sensitivity optical time-domain reflectometer-based distributed intrusion sensor,"
{\it Appl. Opt.}, Vol.~{\bf 53}, 4645, 2015.

\bibitem{Frantz}
Frantz,~L.~M. and Nodvik,~J.~S.
``Theory of pulse propagation in a laser amplifier,"
{\it J. Appl. Phys.}, Vol.~{\bf 34}, 2346, 1963.

\bibitem{Fedorov4}
Zhirnov,~A.~A., Anufriev,~M.~N., Pozhar,~N.~O., Stepanov,~K.~V., Chernutsky,~A.~O., Makhrov,~I.~P., 
Nesterov,~E.~T., Shelestov,~D.~A., Koshelev,~K.~I., Fedorov,~A.~K., Karasik,~V.~E., and Pnev,~A.~B.
``Multipurpose monitoring system for icebreakers: Development, implementation, and testing,"
{\it MATEC Web Conf.}, Vol.~{\bf 75}, 04005, 2016.

\end{thebibliography}
\end{document}